\documentstyle[preprint,eqsecnum,aps,epsfig]{revtex}
\tightenlines
\begin{document}

\draft

\title{BS and DS equations in a Wilson loop context in QCD, effective mass
operator, $q \bar q$ spectrum}

\author{M. Baldicchi and G.M. Prosperi
\footnote{
\sl{Presented by G.M. Prosperi}}}
\address{Dipartimento di Fisica, Universit\`a di Milano\\
I.N.F.N., sezione di Milano\\
via Celoria 16, I20133 Milano, Italy}

\maketitle
\begin{abstract}
\indent
We briefly discuss the quark-antiquark Bethe-Salpeter equation and the quark
Dyson-Schwinger equation derived in preceding papers. We also consider the
$q\bar q$ quadratic mass operator $M^2=(w_1+w_2)^2+U$ obtained by
three-dimensional reduction of the BS equation and the related approximate
center of mass Hamiltonian or linear mass operator $H_{\rm CM}\equiv
M=w_1+w_2+V+\ldots $. We revue previous results on the spectrum and the
Regge trajectories obtained by an approximate diagonalization of $H_{\rm CM}$
and report new results similarly obtained for the original $M^2$. We show that
in both cases we succeed to reproduce fairly well the entire meson spectrum in
the cases in which the numerical calculations were actually practicable and
with the exception of the light pseudoscalar states (related to the chiral
symmetry problematic). A small rearrangement 
of the parameters and the use of a running coupling constant is necessary 
in the $M^2$ case. 
\end{abstract}



\section{Introduction}

In preceding papers \cite{bmp,como} (cf. also \cite{simonov}), using a 
Feynmann-Schwinger type of path integral representation  and an
appropriate ansatz for the Wilson loop correlator, we
have obtained a Bethe-Salpeter and a Dyson-Schwinger equation for certain
``second order'' quark-antiquark and single quark Green functions.

To solve the $q \bar q$ bound state problem one should in principle solve
the DS equation and use the resulting quark propagator in the BS equation.
In practice even an approximate treatment of such problem in its full
four-dimensional form seems to be extremely hard and for we have to
resort to the use of free propagator and of a three-dimensional reduction of
the BS equation (instantaneous approximation). Such reduction takes
the form
of the eigenvalue equation for an effective squared mass operator
$M^2=(w_1+w_2)^2+U$, $w_1$ and $w_2$ being the relativistic free energies of
the quarks and $U$ an interaction related to the BS kernel.

In more conventional terms one can also consider a center of mass Hamiltonian
$H_{\rm CM}\equiv M=w_1+w_2+V+\ldots $ , where at the lowest order $V$ differs
from $U$ only for kinematic factors. This last form can be more directly
compared with usual relativistic and non relativistic potential models and
$V$ turns out to have various significant limit expressions. In the static
limit $V$ takes the Cornell form 
\begin{equation}
V=-{4 \over 3}{\alpha_{\rm s}\over r}+\sigma r \, .
\label{eq:static}
\end{equation}
In the heavy masses limit, by an ${1 \over m}$ expansion (and an appropriate
Foldy-Wouthuysen transformation) it reproduces the semi relativistic
potential discussed in \cite{barchielli} and \cite{como}.  If the spin
dependent terms are neglected, it becomes identical (apart from a question
of ordering) to the potential corresponding to the relativistic flux tube
model \cite{tubolsson}, up to the first order in the coupling
constant $\alpha_{\rm s}$ and the string tension $\sigma$.

In \cite{baldicchi} we have solved numerically the eigenvalue equation for
$H_{\rm CM}$, as we shall explain later, neglecting the
spin-orbit terms but including the hyperfine separation. We have 
succeeded to reproduce fairly well the entire heavy-heavy,
light-light and light-heavy quarkonium spectrum and Regge trajectories
when the actual calculations were feasible. The only real exception was the 
case of the ground light pseudoscalar mesons, for which the 
three-dimensional
reduction of the BS equation does not seem to be appropriate, due to the 
chiral symmetry breaking problem. 

Concerning the choice of the constants, the light quark masses were fixed 
on typical current values, $m_u=m_s=10\, {\rm MeV},\ m_s=200 \, {\rm MeV}$ 
and only the heavy quark masses, the strong coupling constant and the string 
tension were used as fitting parameters. Good agreement with the data was 
found for $m_c=1.40\, {\rm GeV},\
m_b=4.81 \, {\rm GeV}, \ \alpha_{\rm s}=0.363, \ \sigma = 0.175 \, {\rm
GeV^2}$. 

In spite of the success attained it turns out,however,  that the quantity 
$\langle V^2 \rangle$ is not
negligible bringing e.g. to corrections ranging between few tens and 150 MeV
in the $c\bar c$ case. For this reason in this paper we have repeated the
calculations for the more complex operator $M^2$. A good agreement is again 
obtained at the price of a small rearrangement of the parameters and of 
using a running coupling constant given by the usual perturbative expression
\begin{equation}
\alpha_{\rm s}(\bf Q)= {4 \pi \over (11-{2 \over 3}N_{\rm f}) \ln {{\bf Q}^2
      \over \Lambda^2}}
      \label{eq:running}
      \end{equation}
cut at a maximum value $\alpha_{\rm s}(0)$.

    We have taken $N_{\rm f}=4$, $\Lambda=200\, {\rm MeV}$,
$\alpha_{\rm s}(0)=0.35$ and $\sigma = 0.2\, {\rm GeV}^2$, where the last two 
values have been chosen in order to reproduce the correct $J/\Psi - \eta_c$ 
separation and the Regge trajectory slope. We have also chosen 
$m_c=1.394\, {\rm GeV}$ and
$m_b=4.763\, {\rm GeV}$ in order to reproduce exactly the masses of $J/\Psi$
and $\Upsilon$; on the contrary we have left unchanged the masses of the
light quarks.

   In the following we revue the BS and the DS equations in
sect. II and the three-dimensional reduction in sect. III. In sect. IV we
report and discuss the results obtained in \cite{baldicchi} for $H_{\rm CM}$ 
and the new results for $M^2$.


\section{Bethe-Salpeter and Dyson-Schwinger equations}

The gauge invariant ``second order'' Green functions considered in 
\cite{bmp}, \cite{como} and \cite{simonov} were defined as
\begin{equation}
H^{\rm gi}(x_1,x_2;y_1,y_2) = -{1\over 3} {\rm Tr _C}
\langle U(x_2,x_1) \Delta_1^\sigma (x_1,y_1;A)
 U(y_1,y_2) \tilde{\Delta}_2^\sigma (x_2,y_2;-\tilde{A})\rangle \,,
 \label{eq:qqeqh}
 \end{equation}
 \begin{equation}
 H^{\rm gi}(x-y) = i {\rm Tr _C}
 \langle U(y,x) \Delta ^\sigma (x,y;A) \rangle \, ,
 \label{eq:qeqh} 
 \end{equation}
 where the tilde and  ${\rm Tr_C}$ denote the transposition and the
 trace, respectively, over the color indices alone; $U$ is a path-ordered
 gauge string joining $a$ to $b$ (Schwinger string),
 $
 U(b,a)= {\rm P}  \exp \left \{ig\int_a^b dx^{\mu} \, A_{\mu}(x)
 \right \}
 $;
 while $\Delta^\sigma(x,y;A)$ stands for the ``second order'' quark propagator
 in a external gauge field $A^\mu$, defined by the iterated Dirac equation
 \begin{equation}
 (D_\mu D^\mu +m^2 -{1\over 2} g \, \sigma^{\mu \nu} F_{\mu \nu})
 \Delta^\sigma (x,y;A) = -\delta^4(x-y) \, ,
 \label{eq:propk}
 \end{equation}
with $\sigma^{\mu \nu} = {i\over 2} [\gamma^\mu, \gamma^\nu]$; finally
the angle brackets in (\ref{eq:qqeqh}) and (\ref {eq:qeqh}) denote average 
on the gauge variables alone (weighted in principle
with the determinant $M_f(A)$ resulting from the explicit integration of
the fermionic fields).

The quantities $H^{\rm gi}(x_1,x_2;y_1,y_2)$ and  $H^{\rm gi}(x-y)$ are
simply related to their ordinary ``first order'' counterparts
and can be equivalently used for the determination of the bound state. The 
advantage they offer is that of admitting path integral 
representations in terms of quark world lines which derive from the similar 
Feynmann-Schwinger representation for $\Delta^\sigma(x,y;A)$. 
Such representations depend on the gauge field only trough Wilson 
correlators
$
   W[\Gamma] = {1\over 3}
      \langle {\rm Tr} {\rm P} \exp \{ i g  \oint_\Gamma dx^{\mu}
      A_{\mu} \}  \rangle  \,,
$
associated to loops $\Gamma$ made by the quark and antiquark world
lines closed by ``Schwinger strings''. 

In principle, as a consequence, the above correlators should
determine the whole dynamics. Unfortunately, due to confinement and the
consequent failure of a purely perturbative approach, a consistent analytic
evaluation of $W$ from the Lagrangian alone is not possible today and one
has to rely on models based on incomplete theoretical arguments and 
lattice simulation information. The most naive, but at the same time less 
arbitrary assumption, consists
in writing $i\ln W$ as the sum of its perturbative expression and an
area term (modified area law (MAL) model)
  \begin{equation}
  i \ln W = i (\ln W)_{\rm pert} + \sigma S_{\rm min}   \,  ,
  \label{eq:wl}
  \end{equation}
where the first quantity is supposed to give correctly the short range
limit, the second the long range one. Notice in principle any more
sophisticated model could be used, at the condition it preserves
certain general properties of functional derivability of the exact
expression. In practice not even (\ref{eq:wl}) can be
treated exactly. Actually one has to replace the minimal surface
$S_{\rm min}$ by its ``equal time straight line approximation'', defined as
the surface spanned by straight lines joining equal time opposite points of
the loop $\Gamma$ in the $q \bar q$ center of mass frame.

  The path integral representations obtained in such a way could be
used directly for numerical calculations or for analytic developments. In the
last context it is convenient to consider a second type of second order
functions $H(x_1,x_2,y_1,y_2)$ and $H(x-y)$ obtained from
$H^{\rm gi}(x_1,x_2,y_1,y_2)$ and $H^{\rm gi}(x-y)$
by omitting in their path integral representation the contributions
to $i \ln W$ coming from gluon lines or straight lines involving points
of the Schwinger strings. In the limit of vanishing $x_1-x_2,\
y_1-y_2$ or $x-y$ such new quantities coincide with the original ones and
are completely equivalent to them for what concerns the determination of  
bound states, effective masses, quark condensates, etc. 

By an appropriate recurrence method, an inhomogeneous Bethe-Salpeter
equation and a Dyson-Schwinger equation can be derived for 
$H(x_1,x_2,y_1,y_2)$ and $H(x-y)$, respectively. In the momentum space, 
the corresponding homogeneous BS-equation can be written (in a $4 \times 4$ 
matrix representation)
\begin{eqnarray}
  \Phi_P (k) &=& -i \int {d^4u \over (2 \pi)^4}
        \hat I_{ab} \big (k-u, {1 \over 2}P +{k+u \over 2},
              {1 \over 2}P-{k+u \over 2} \big )\nonumber \\
   & & \qquad \qquad \qquad
     \hat H_1   \big ({1 \over 2} P  + k \big )\sigma^a  \Phi_P (u) \sigma^b
     \hat H_2 \big (-{1 \over 2} P + k \big ) \, ,
\label{eq:bshoma}
\end{eqnarray}
where $\Phi_P (k)$ denotes an appropriate wave function and the center of
mass frame has to be understood; i.e. $P=(m_B, {\bf 0})$, $m_B$ being the
bound state mass. Similarly, in terms of the irreducible self-energy 
defined by
$\hat H(k) =\hat H_0(k) + i\hat H_0(k)\hat \Gamma (k) \hat H(k) \,$
the DS-equation can be written also
\begin{equation}
\hat \Gamma(k) =  \int {d^4 l \over (2 \pi)^4}  \,
\hat I_{ab} \Big ( k-l;{k+l \over 2},{k+l \over 2} \Big )
\sigma^a \hat H(l) \, \sigma^b \ .
\label{eq:sdeq}
\end{equation}

  Notice that in principle  (\ref{eq:bshoma}) and  (\ref{eq:sdeq}) are 
exact equations. However
the kernels $\hat I_{ab}$ are generated in the form of an expansion in
$\alpha_{\rm s}$ and $\sigma$. At the lowest order in both such
constants, we have explicitly
\begin{eqnarray}
& & \hat I_{0;0} (Q; p, p^\prime)  = 
       16 \pi {4 \over 3} \alpha_{\rm s} p^\alpha p^{\prime \beta}
        \hat D_{\alpha \beta} (Q)  + \nonumber \\
& &  \qquad \qquad + 4 \sigma  \int d^3 {\bf \zeta} e^{-i{\bf Q}
              \cdot {\bf \zeta}}
         \vert {\bf \zeta} \vert \epsilon (p_0) \epsilon (p_0^\prime )
         \int_0^1 d \lambda \{ p_0^2 p_0^{\prime 2} -
         [\lambda p_0^\prime {\bf p}_{\rm T} +
         (1-\lambda) p_0 {\bf p}_{\rm T}^\prime ]^2 \} ^{1 \over 2}
\nonumber \\
& & \hat I_{\mu \nu ; 0}(Q;p,p^\prime) = 4\pi i {4 \over 3} \alpha_{\rm s}
   (\delta_\mu^\alpha Q_\nu - \delta_\nu^\alpha Q_\mu) p_\beta^\prime
   \hat D_{\alpha \beta}(Q)  - \nonumber \\
& & \qquad \qquad \qquad  - \sigma  \int d^3 {\bf \zeta} \, e^{-i {\bf
      Q} \cdot \zeta} \epsilon (p_0) {\zeta_\mu p_\nu -\zeta_\nu p_\mu \over
      \vert {\bf \zeta} \vert \sqrt{p_0^2-{\bf p}_{\rm T}^2}}
      p_0^\prime  \nonumber \\
& & \hat I_{0; \rho \sigma}(Q;p,p^\prime) =
     -4 \pi i{4 \over 3} \alpha_{\rm s}
     p^\alpha (\delta_\rho^\beta Q_\sigma - \delta_\sigma^\beta Q_\rho)
     \hat D_{\alpha \beta}(Q) + \nonumber \\
& & \qquad \qquad  \qquad  + \sigma  \int d^3 {\bf
\zeta} \, e^{-i{\bf Q}
  \cdot {\bf \zeta}} p_0
  {\zeta_\rho p_\sigma^\prime - \zeta_\sigma p_\rho^\prime \over
    \vert {\bf \zeta} \vert \sqrt{p_0^{\prime 2}
   -{\bf p}_{\rm T}^{\prime 2}} }
    \epsilon (p_0^\prime)  \nonumber \\
& & \hat I_{\mu \nu ; \rho \sigma}(Q;p,p^\prime) =
    \pi {4\over 3} \alpha_{\rm s}
    (\delta_\mu^\alpha Q_\nu - \delta_\nu^\alpha Q_\mu)
    (\delta_\rho^\alpha Q_\sigma - \delta_\sigma^\alpha Q_\rho)
     \hat D_{\alpha \beta}(Q)
 \label{eq:imom}
 \end{eqnarray}
\noindent
where in the second and in the third equation $\zeta_0 = 0$ has to be
understood. Notice that the use of (\ref{eq:running}) in (\ref{eq:imom}) would
amount to include higher order contributions.


\section{Three-dimensional reduction of the BS equation}

To obtain from (\ref{eq:bshoma}) a three-dimensional equation we can perform
on such equation the so called instantaneous approximation. This consists
in replacing in (\ref{eq:bshoma}) $\hat{H}_{2}^{(j)}(p)$ with the free 
quark propagator ${-i \over p^2 -m^2}$ and the kernel 
$\hat{I}_{ab}$ with $ \hat{I}_{ab}^{\rm inst}({\bf k}, {\bf k}^\prime)$ 
obtained from $\hat{I}_{ab}$
setting $k_0=k_0^\prime= \eta_2 { w_1+ w_1^\prime \over 2}-
\eta_1 {w_2 +w_2^\prime \over 2}$ with $w_j = \sqrt{m_j^2 + {\bf k}^2} $
and $ w_j^\prime = \sqrt{m_j^2 + {\bf k}^{\prime 2}}$. Then, by performing
explicitly the integration over $k_0^{\prime}$ and further integrating 
the resulting expression in $k_0$, we obtain 
\begin{eqnarray}
  & &  (w_1 + w_2 )^2 \varphi_{m_B} ({\bf
          k}) +\nonumber \\
  & & \quad + \int {d^3 k^\prime  \over (2 \pi )^3 }
           \sqrt{ w_1 + w_2  \over 2
            w_1 w_2} \hat{I}_{ab}^{\rm inst}({\bf k},
             {\bf k}^\prime)  \sqrt{w_1^\prime + w_2 ^\prime
             \over 2 w_1^\prime w_2^\prime }
            \sigma^a \varphi_{m_B}({\bf k}^\prime) \sigma^b 
            = m_B^2 \varphi_{m_B}({\bf k}) \,,
\label{eq:cinqqua}
\end{eqnarray}
with
$   \varphi_{P} ({\bf k}) =
     \sqrt{ 2 w_1 w_2 \over w_1+w_2}
     \int_{-\infty}^\infty d k_0 \Phi_P ({ k}).
$

Eq. (\ref{eq:cinqqua}) is the eigenvalue equation  for the squared mass
operator,
\begin{equation}
    M^2 = M_0^2 + U
\label{eq:quadr}
\end{equation}
with  $ M_0 = \sqrt{m_1^2 + {\bf k}^2} + \sqrt{m_2^2 + {\bf k}^2} $ and
\begin{equation}
   \langle {\bf k} \vert U \vert {\bf k}^\prime \rangle =
        {1\over (2 \pi)^3 }
        \sqrt{ w_1 + w_2 \over 2  w_1  w_2}\, \hat I_{ab}^{\rm inst}
        ({\bf k} , {\bf k}^\prime)  \sqrt{ w_1^\prime + w_2^\prime \over 2
         w_1^\prime w_2^\prime}\sigma_1^a \sigma_2^b \,.
\label{eq:quadrrel}
\end{equation}
The quadratic form of Eq.(\ref{eq:quadr}) obviously derives from the second 
order character of the formalism we have used. 

  In more usual terms one can also write
\begin{equation}
          H_{\rm CM} \equiv M = M_0 + V + \ldots 
\label{eq:lin}
\end{equation}
with
\begin{equation}
\langle {\bf k} \vert V \vert {\bf k}^\prime \rangle =
   {1\over w_1 + w_2 + w_1^\prime + w_2^\prime }
   \langle {\bf k} \vert U \vert {\bf k}^\prime \rangle
   ={1 \over ( 2 \pi)^3 }
    {1 \over 4 \sqrt{ w_1 w_2 w_1^\prime w_2^\prime } } \hat{I}_{ab}^{\rm inst}
    ({\bf k}, {\bf k}^\prime)\sigma_1^a \sigma_2^b 
\label{eq:linrel}
\end{equation}
In (\ref{eq:lin}) the dots stand for higher order terms in $\alpha _{\rm s}$
and $\sigma$ and kinematic factors equal to 1 on the energy shell
have been neglected.

   From Eqs. (\ref{eq:quadrrel}) and (\ref{eq:imom}) one obtains explicitly
  \begin{eqnarray}
   & & \langle {\bf k} \vert U \vert {\bf k}^\prime \rangle =
      \sqrt{(w_1+w_2) (w_1^\prime
     +w_2^\prime)
     \over w_1 w_2 w_1^\prime w_2^\prime} \Big \{ 
     \bigg [ - {4\over 3} {\alpha_s \over \pi^2}{1\over {\bf Q}^2}
     \big [ q_{10} q_{20} + {\bf q}^2 - { ( {\bf Q}\cdot {\bf q})^2 \over
     {\bf Q}^2 } ] \nonumber \\
& & + {i\over 2 {\bf Q}^2} {\bf k}\times {\bf k}^\prime \cdot ({\bf
     \sigma}_1 + {\bf \sigma}_2 ) + {1\over 2 {\bf Q}^2 } [ q_{20}
     (\alpha_1 \cdot {\bf Q}) - q_{10} (\alpha_2\cdot {\bf Q}) ]+
    \nonumber \\
& & + {1\over 6} {\bf \sigma}_1 \cdot {\bf \sigma}_2 + {1\over 4}
    \left ( {1\over 3} {\bf \sigma}_1 \cdot {\bf \sigma}_2 - 
    { ( {\bf Q}\cdot \sigma_1)
    ( {\bf Q}\cdot {\bf \sigma_2}) \over {\bf Q}^2 } \right )
    + {1\over 4 {\bf Q}^2 } ( \alpha_1 \cdot {\bf Q}) ( \alpha_2 \cdot {\bf Q})
    \bigg ] + \nonumber \\ 
& & +{1\over ( 2 \pi)^3} \int d^3{\bf r} e^{i {\bf Q}\cdot {\bf r}}
J^{\rm inst}({\bf r}, {\bf q}, q_{10}, q_{20})  \Big \}
\label{eq:upot}
\end{eqnarray}
with
\begin{eqnarray}
& &
J^{\rm inst}({\bf r}, {\bf q}, q_{10}, q_{20})= {\sigma r \over q_{10}+q_{20}}
   [ q_{20}^2 \sqrt{q_{10}^2-{\bf q}^2_t} + q_{10}^2 \sqrt{q_{20}
   -{\bf q}_{\rm T}^2}] +\nonumber\\
& & \quad + {q_{10}^2 q_{20}^2 \over \vert {\bf q}_{\rm T} \vert}
   (\arcsin{\vert{\bf q}_{\rm T}\vert \over q_{10} }
   + \arcsin{\vert {\bf q}_{\rm T}\vert \over q_{20}} )]
   \label{eq:uconf1}\\
& & \quad - {\sigma\over r} \left [ {q_{20} \over 
    \sqrt{q_{10}^2-{\bf q}^2_{\rm T}}}
   ( {\bf r} \times {\bf q}\cdot \sigma_1 + i q_{10} ({\bf r}\cdot \alpha_1))
   + {q_{10} \over \sqrt{q_{20}^2 - {\bf q}^2_{\rm T}}}
   ( {\bf r}\times {\bf q} \cdot
   \sigma_2 - i q_{20} ( {\bf r}\cdot{\bf \alpha}_2)) \right ] \nonumber
\end{eqnarray}
Here $ \alpha_j^k $ denote the usual Dirac matrices $\gamma_j^0 \gamma_j^k$,
$\sigma_j^k$ the $4 \times 4$ Pauli matrices $ \left( \matrix {\sigma_j^k &
0 \cr 0 & \sigma_j^k}\right ) $ and obviously
${\bf q} ={ {\bf k}+ {\bf k}^\prime \over 2}\,, \quad {\bf Q}= {\bf k}
- {\bf k}^\prime \,, \quad q_{j0}= {w_j+w_j^\prime \over 2}$. \
Notice that, due to the terms in $ \alpha_j^k $, such $U$ is self adjoint only
with reference to the undefined metric operator $\gamma_1^0 \gamma_2^0$.

   Due to (\ref{eq:linrel}) the potential $V$ can be obtained from $U$ as 
given by (\ref{eq:upot}-\ref{eq:uconf1}) simply by the kinematic replacement
$
\sqrt{ (w_1+w_2) (w_1^\prime +w_2^\prime)\over w_1w_2w_1^\prime w_2^\prime}
\to {1\over 2\sqrt{w_1 w_2 w_1^\prime w_2^\prime}}
$.


\section{Determination of the spectrum}

\indent
   In ref. \cite{baldicchi} we have evaluated the eigenvalues of the
operator $H_{\rm CM}$ for the potential $V$ discussed above omitting the
spin-orbit
terms and including only the hyperfine splitting. The numerical procedure we
have followed is very simple. It consists in solving first the
eigenvalue equation for the static potential (\ref{eq:static}) by
the Rayleigh-Ritz method \cite{simon} using the three-dimensional harmonic
oscillator basis diagonalizing a $30 \times 30$ matrix. Then we have
evaluated the quantities $\langle \psi_{\nu}|H_{\rm CM}|\psi_{\nu} \rangle$
for the
eigenfunctions $\psi_{\nu}$ obtained in the first step, choosing
the scale parameter occurring in the basis in order to make minimum the
ground state mass  $\langle \psi_{1S}| H_{\rm CM}|\psi_{1S} \rangle$.
Notice that the determination of $\langle \psi_{\nu} |V|\psi_{\nu} \rangle$ 
for the exact $V$ is not trivial, since in general one should evaluate
five-dimensional integrals of a highly singular functions. For such reason we 
have used two different expansions for high and low transversal momentum 
(angular momentum), that allows to reduce to a three-dimensional integrals 
and treated the singularity with the method suggested in \cite{maung}. 

The procedure we have followed for the determination of the eigenvalues of
$M^2$ is essentially the same. Again we solve first the eigenvalue equation
for $H_{\rm CM}$ with the static potential, and then we evaluate the 
quantities $\langle \psi_{\nu} |M^2|\psi_{\nu} \rangle$. In this case the 
hyperfine splitting is determined by the equation
\begin{eqnarray}
& & (^3M_{nl})^2-(^1M_{nl})^2 = \nonumber \\
& & \quad     {32 \over 9\pi} \int_0^\infty \!
     dk \, k^2  \int_0^\infty \! dk^{\prime} \,
     k^{\prime 2} \Psi_{nl}^* (k)  \Psi_{nl} (k^{\prime}) 
     \sqrt {{w_1+w_2 \over w_1 w_2}} \sqrt {{w_1^{\prime}+w_2^{\prime}
     \over w_1^{\prime} w_2^{\prime}}} \int_{-1}^1 \! d\xi \,
     \alpha_{\rm s}({\bf Q}) P_l(\xi)\, ,
\label{eq:hyper}
\end{eqnarray}
which is more complicated than the corresponding equation in the case of
$H_{\rm CM}$ \cite{baldicchi}.

\begin{figure}[htbp!]
  \begin{center}
    \leavevmode
    \setlength{\unitlength}{1.0mm}
    \begin{picture}(140,70)
      \put(-10,0){\mbox{\epsfig{file=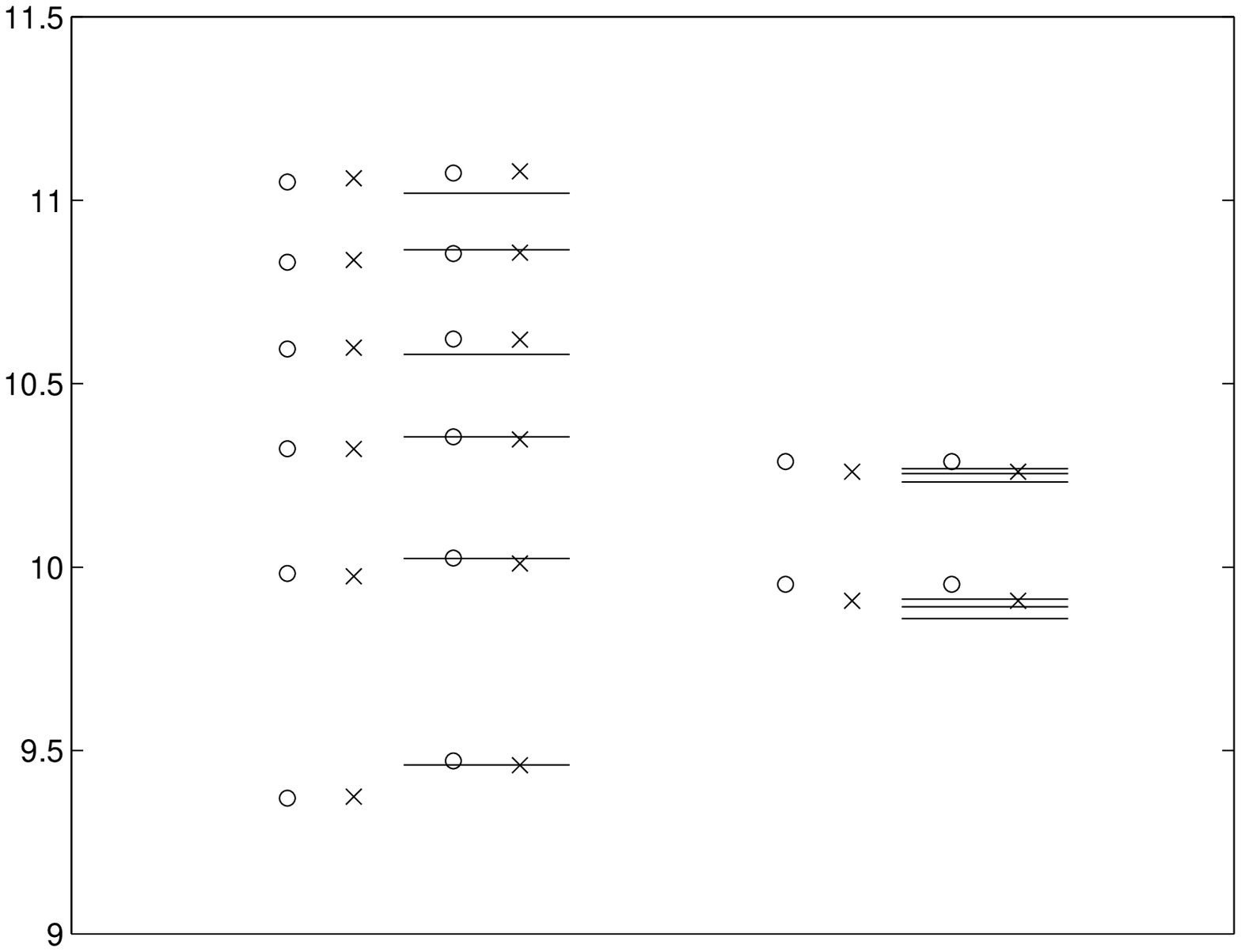,height=7cm,width=7.5cm}}}
      \put(75,0){\mbox{\epsfig{file=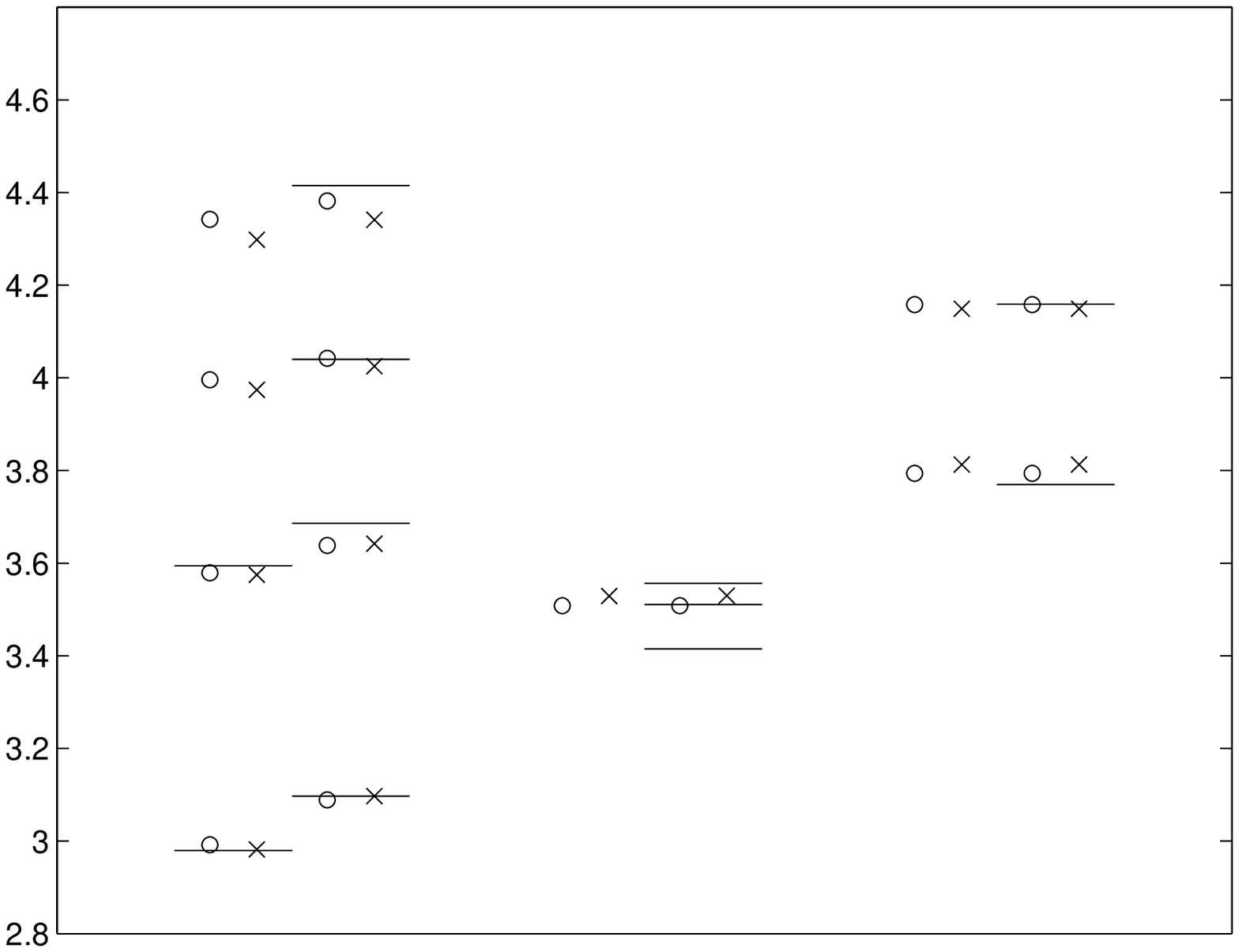,height=7cm,width=7.5cm}}}
      \put(46,5){ $ b \bar{b} $ }
      \put(130,5){ $ c \bar{c} $ }
      \put(5,61){ $ {^{1} {\rm S}_{0}} $ }
      \put(15,61){ $ {^{3} {\rm S}_{1}} $ }
      \put(35,61){ $ {^{1} {\rm P}_{1}} $ }
      \put(45,61){ $ {^{3} {\rm P}_{J}} $ }
      \put(84,61){ $ {^{1} {\rm S}_{0}} $ }
      \put(91,61){ $ {^{3} {\rm S}_{1}} $ }
      \put(105,61){ $ {^{1} {\rm P}_{1}} $ }
      \put(112,61){ $ {^{3} {\rm P}_{J}} $ }
      \put(127,61){ $ {^{1} {\rm D}_{1}} $ }
      \put(134,61){ $ {^{3} {\rm D}_{J}} $ }
      \put(25,61){ n }
      \put(25,55){ $ {}_{6} $ }
      \put(25,50){ $ {}_{5} $ }
      \put(25,43){ $ {}_{4} $ }
      \put(25,37){ $ {}_{3} $ }
      \put(25,28){ $ {}_{2} $ }
      \put(25,13){ $ {}_{1} $ }
      \put(55,61){ n }
      \put(55,35){ $ {}_{2} $ }
      \put(55,25){ $ {}_{1} $ }
      \put(100,61){ n }
      \put(100,55){ $ {}_{4} $ }
      \put(100,43){ $ {}_{3} $ }
      \put(100,30){ $ {}_{2} $ }
      \put(100,9){ $ {}_{1} $ }
      \put(121,61){ n }
      \put(121,25){ $ {}_{1} $ }
      \put(143,55){ $ {}_{J} $ }
      \put(143,47){ $ {}_{1} $ }
      \put(143,34){ $ {}_{1} $ }
    \end{picture}
  \end{center}
\caption{Heavy-heavy quarkonium spectra}
\label{fig1}
\end{figure}

\begin{figure}[htbp!]
  \begin{center}
    \leavevmode
    \setlength{\unitlength}{1.0mm}
    \begin{picture}(140,70)
      \put(-11,0){\mbox{\epsfig{file=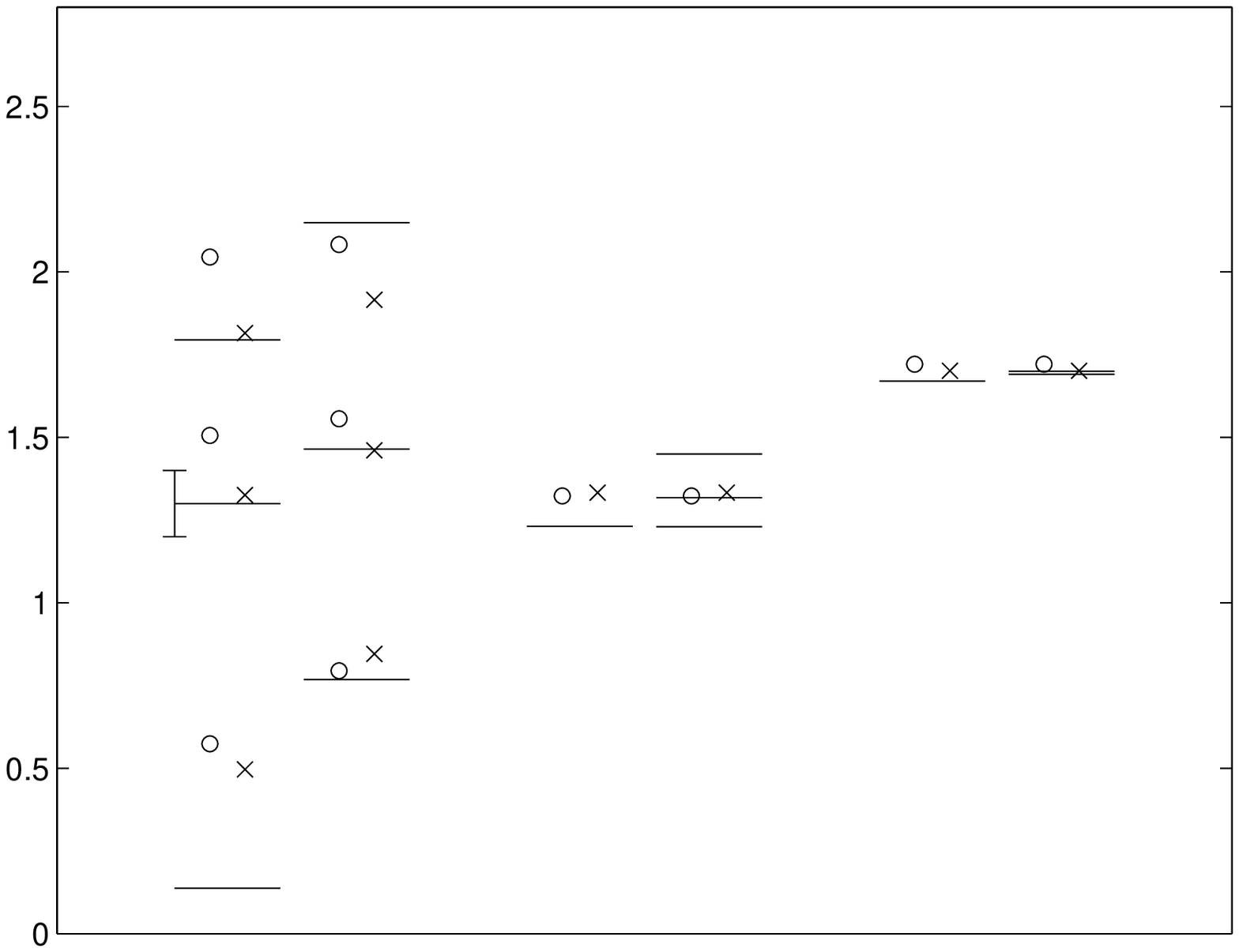,height=7cm,width=5.5cm}}}
      \put(45,0){\mbox{\epsfig{file=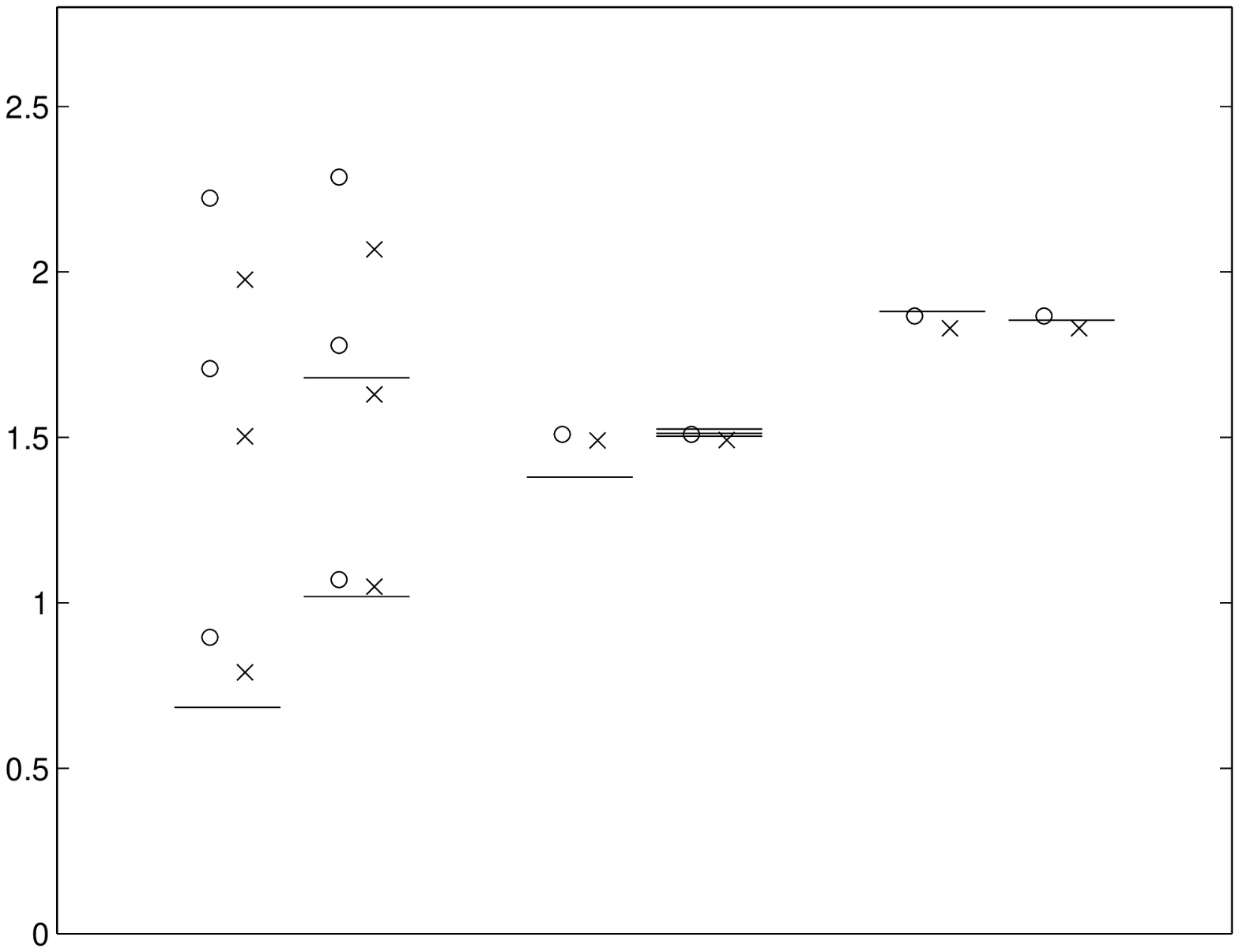,height=7cm,width=5.5cm}}}
      \put(101,0){\mbox{\epsfig{file=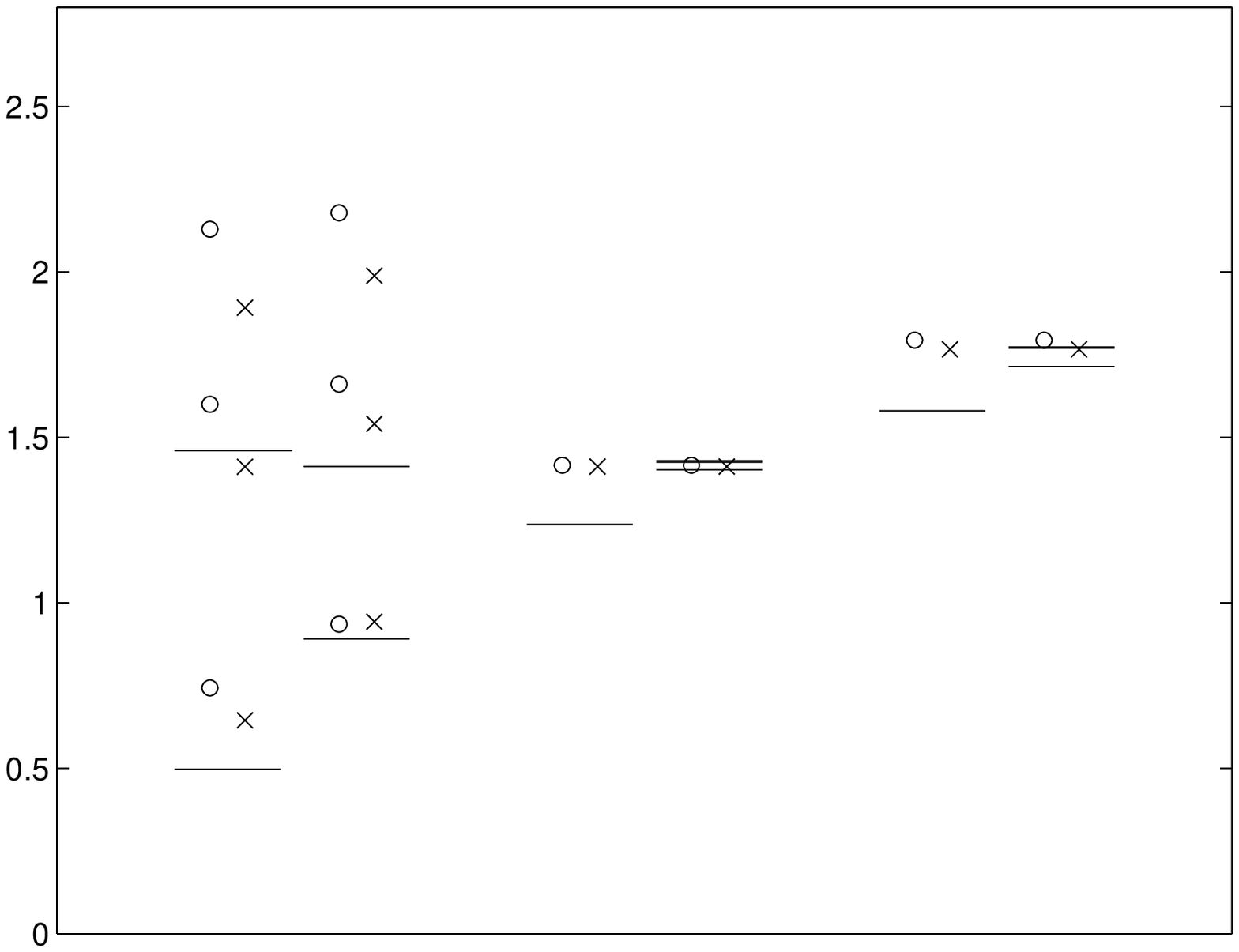,height=7cm,width=5.5cm}}}
      \put(30,5){ $ u \bar{u} $ }
      \put(-7,61){ $ {^{1} {\rm S}_{0}} $ }
      \put(0,61){ $ {^{3} {\rm S}_{1}} $ }
      \put(9,61){ $ {^{1} {\rm P}_{1}} $ }
      \put(16,61){ $ {^{3} {\rm P}_{J}} $ }
      \put(25,61){ $ {^{1} {\rm D}_{1}} $ }
      \put(32,61){ $ {^{3} {\rm D}_{J}} $ }
      \put(39,55){ $ {}_{J} $ }
      \put(37.5,42.5){ $ {}_{1,3} $ }
      \put(22,36.5){ $ {}_{?} $ }
      \put(85,5){ $ s \bar{s} $ }
      \put(49,61){ $ {^{1} {\rm S}_{0}} $ }
      \put(56,61){ $ {^{3} {\rm S}_{1}} $ }
      \put(65,61){ $ {^{1} {\rm P}_{1}} $ }
      \put(72,61){ $ {^{3} {\rm P}_{J}} $ }
      \put(81,61){ $ {^{1} {\rm D}_{1}} $ }
      \put(88,61){ $ {^{3} {\rm D}_{J}} $ }
      \put(95,55){ $ {}_{J} $ }
      \put(95,46){ $ {}_{3} $ }
      \put(72,34.5){ $ {}_{?} $ }
      \put(57,16.5){ $ \eta_{S} $ }
      \put(140,5){ $ u \bar{s} $ }
      \put(106,61){ $ {^{1} {\rm S}_{0}} $ }
      \put(113,61){ $ {^{3} {\rm S}_{1}} $ }
      \put(122,61){ $ {^{1} {\rm P}_{1}} $ }
      \put(129,61){ $ {^{3} {\rm P}_{J}} $ }
      \put(138,61){ $ {^{1} {\rm D}_{1}} $ }
      \put(143,61){ $ {^{3} {\rm D}_{J}} $ }
      \put(150,55){ $ {}_{J} $ }
      \put(149.5,45){ $ {}_{2,3} $ }
      \put(150,42){ $ {}_{1} $ }
      \put(128,31){ $ {}_{?} $ }
      \put(143.5,39.5){ $ {}_{?} $ }
    \end{picture}
  \end{center}
\caption{Light-light quarkonium spectra. Question marks indicate
states whose assignment to a multiplet is not obvious.
Position of $ \eta_{S} $ is derived from $ \eta $ and
$ \eta^{\prime}$, by standard mixture assumption}
\label{fig2}
\end{figure}

\begin{figure}[htbp!]
  \begin{center}
    \leavevmode
    \setlength{\unitlength}{1.0mm}
    \begin{picture}(140,70)
      \put(-10,0){\mbox{\epsfig{file=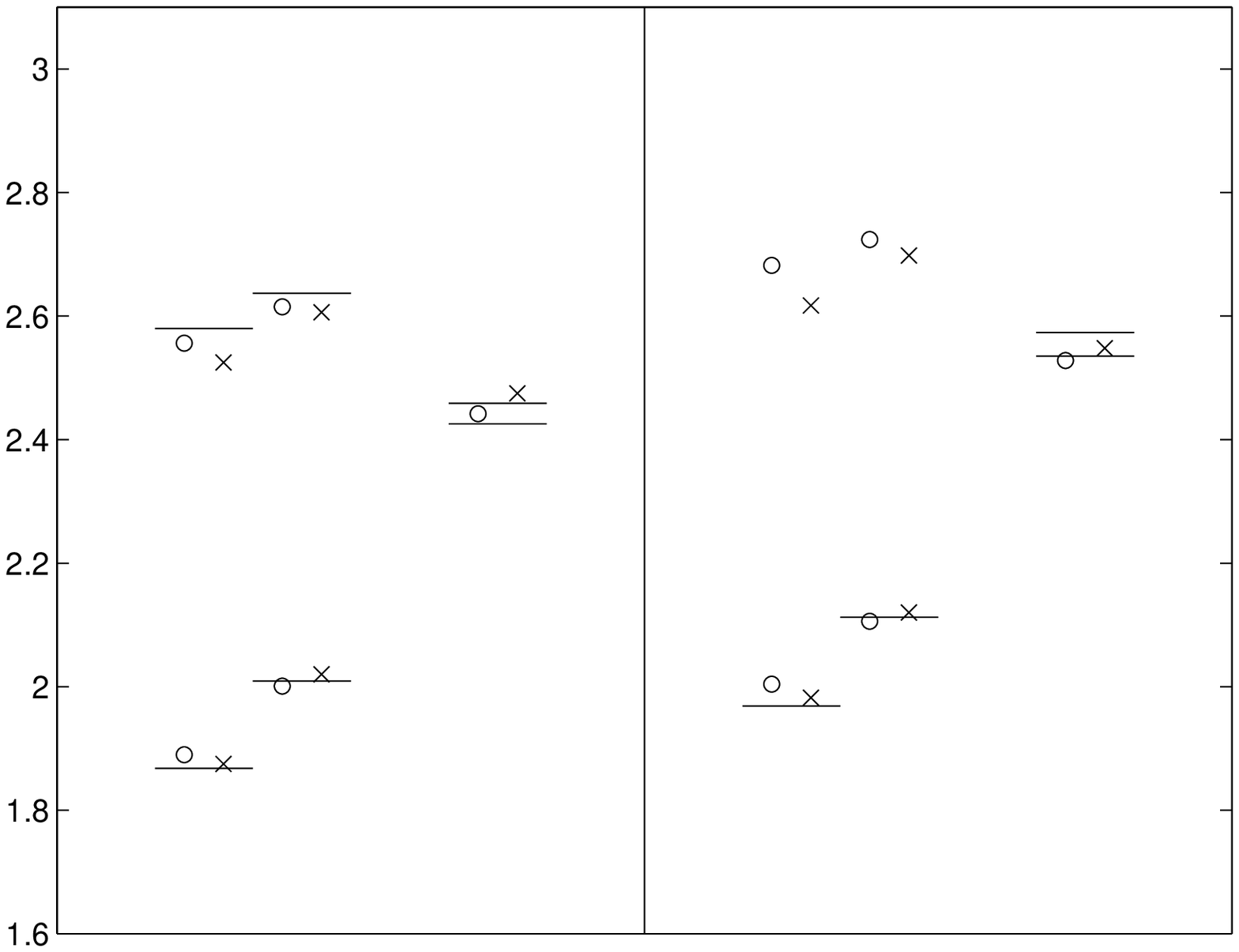,height=7cm,width=7.5cm}}}
      \put(75,0){\mbox{\epsfig{file=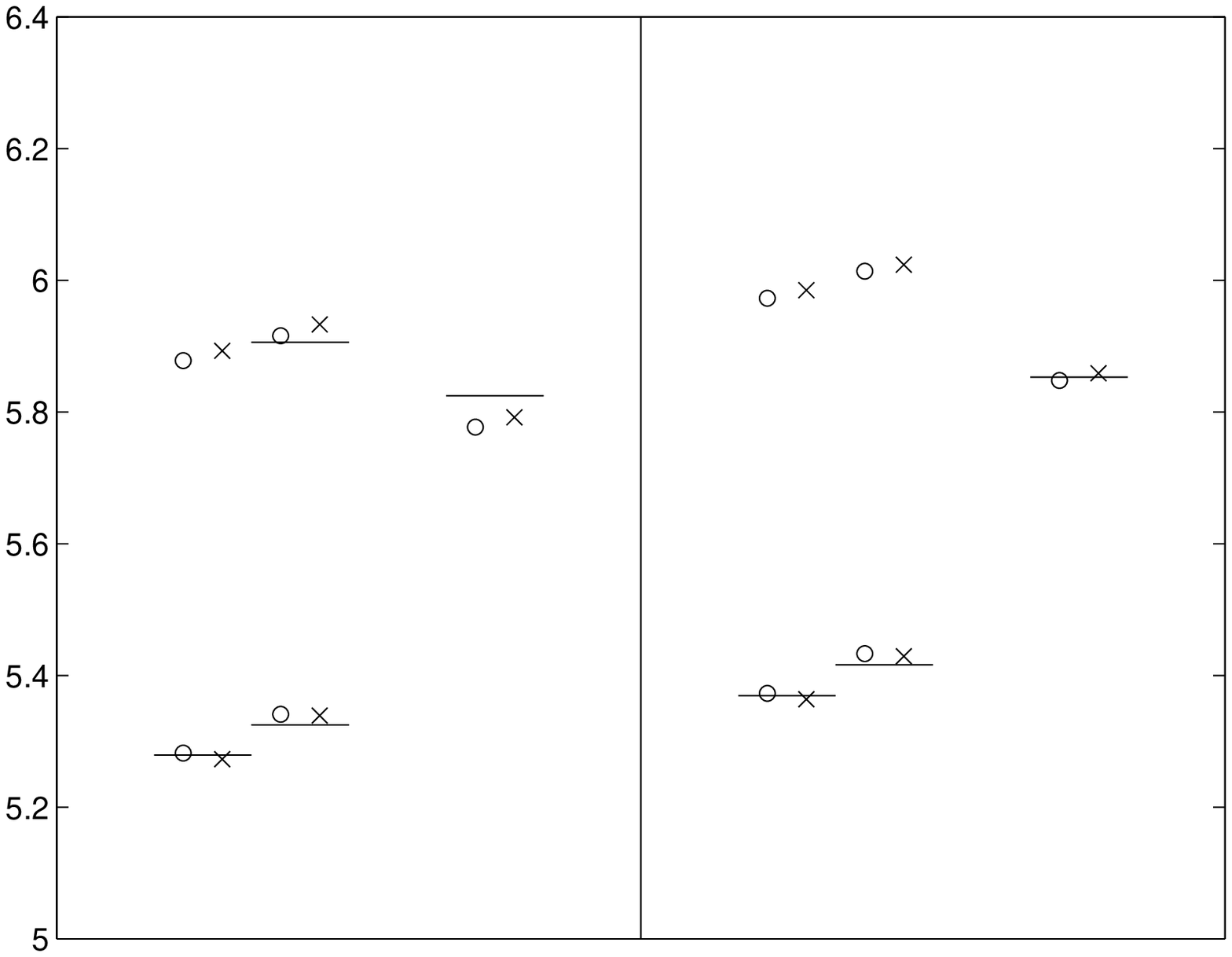,height=7cm,width=7.5cm}}}
      \put(17,5){ $ u \bar{c} $ }
      \put(53,5){ $ s \bar{c} $ }
      \put(-3,61){ $ {^{1} {\rm S}_{0}} $ }
      \put(4,61){ $ {^{3} {\rm S}_{1}} $ }
      \put(16,61){ $ {^{3} {\rm P}_{J}} $ }
      \put(23,55){ $ {}_{J} $ }
      \put(23,41){ $ {}_{2} $ }
      \put(23,38){ $ {}_{1} $ }
      \put(32,61){ $ {^{1} {\rm S}_{0}} $ }
      \put(39,61){ $ {^{3} {\rm S}_{1}} $ }
      \put(51,61){ $ {^{3} {\rm P}_{J}} $ }
      \put(59,55){ $ {}_{J} $ }
      \put(59,46){ $ {}_{2} $ }
      \put(59,43){ $ {}_{1} $ }
      \put(102,5){ $ u \bar{b} $ }
      \put(138,5){ $ s \bar{b} $ }
      \put(82,61){ $ {^{1} {\rm S}_{0}} $ }
      \put(89,61){ $ {^{3} {\rm S}_{1}} $ }
      \put(101,61){ $ {^{3} {\rm P}_{J}} $ }
      \put(117,61){ $ {^{1} {\rm S}_{0}} $ }
      \put(124,61){ $ {^{3} {\rm S}_{1}} $ }
      \put(136,61){ $ {^{3} {\rm P}_{J}} $ }
     \end{picture}
  \end{center}
\caption{light-heavy quarkonium spectra}
\label{fig3}
\end{figure}

\begin{figure}[htbp!]
  \begin{center}
    \leavevmode
    \setlength{\unitlength}{1.0mm}
    \begin{picture}(140,70)
      \put(-10,0){\mbox{\epsfig{file=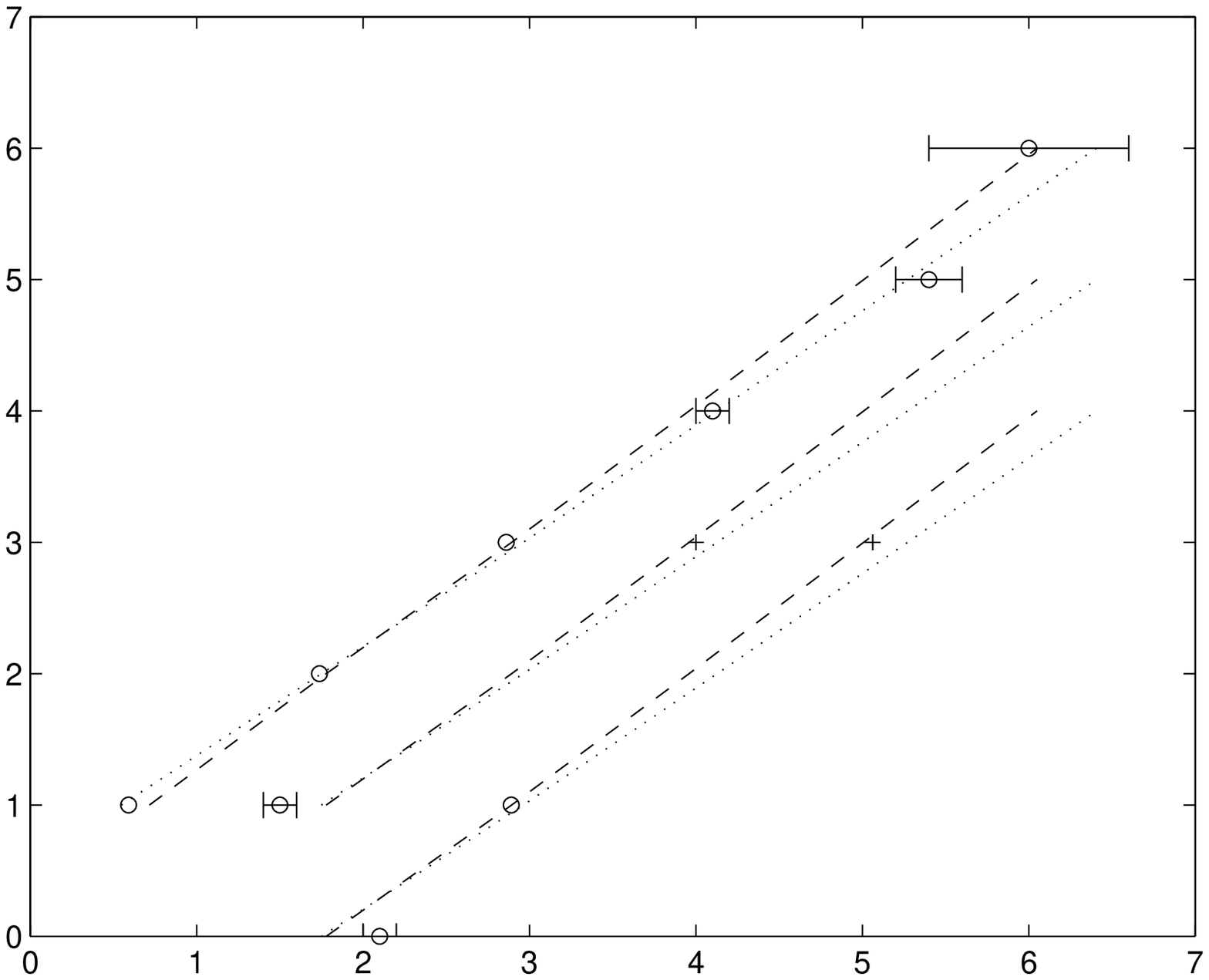,height=7cm,width=7cm}}}
      \put(78,0){\mbox{\epsfig{file=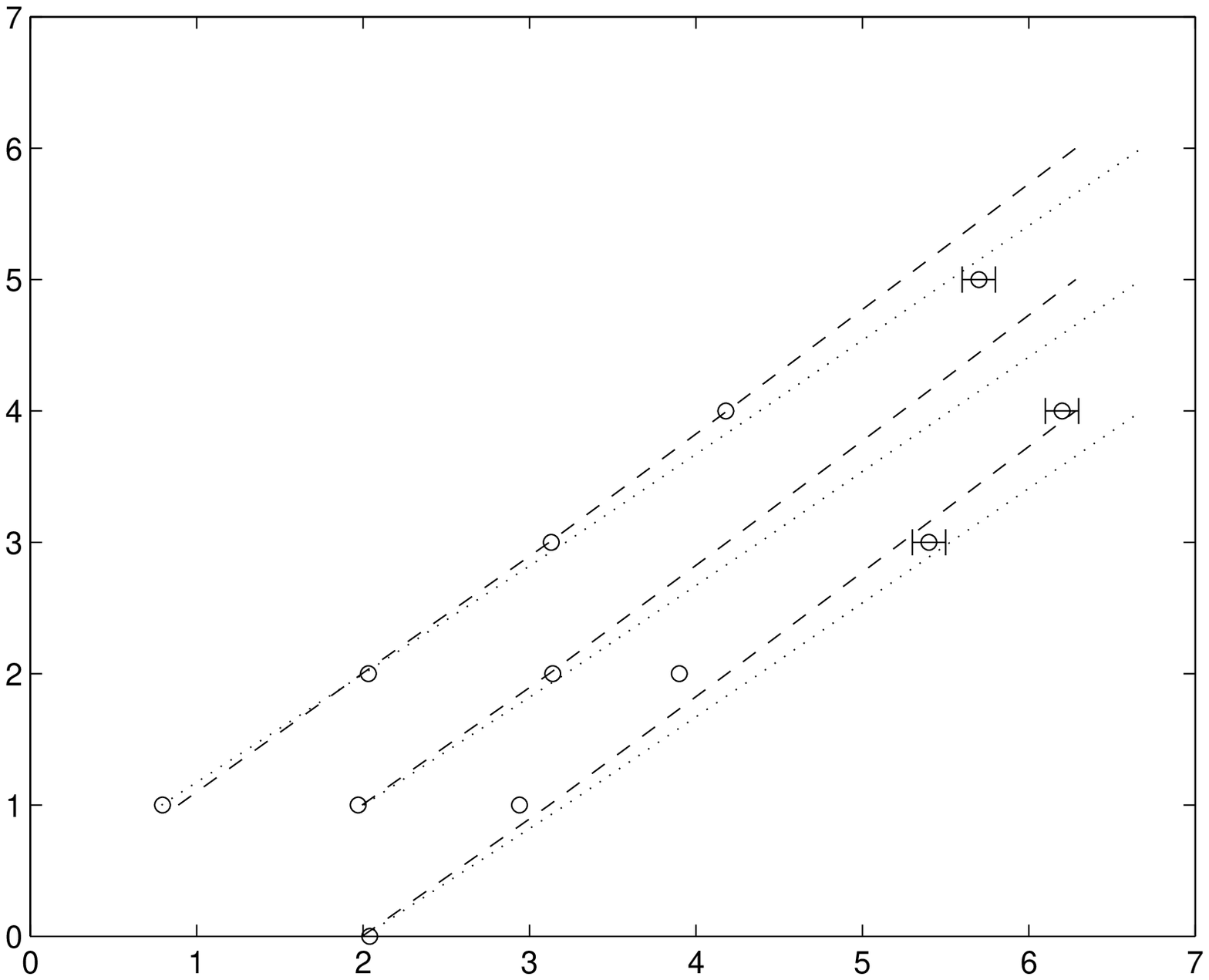,height=7cm,width=7cm}}}
      \put(1,60){ $ u \bar{u} $ }
      \put(-14,63){ $ J $ }
      \put(60,1){ $ M^{2} $ }
      \put(50,61){ $ J = L + 1 $ }
      \put(50,52){ $ J = L $ }
      \put(50,43){ $ J = L - 1 $ }
      \put(-16,14){ $ \rho ( 770 ) $ }
      \put(-9,22){ $ a_{2} ( 1320 ) $ }
      \put(-8,7){ $ a_{1} ( 1260 ) $ }
      \put(4,-3){ $ a_{0} ( 1450 ) $ }
      \put(1,31){ $ \rho_{3} ( 1690 ) $ }
      \put(19,11){ $ \rho ( 1700 ) $ }
      \put(12,40){ $ a_{4} ( 2040 ) $ }
      \put(18,26){ $ X ( 2000 ) $ }
      \put(22,50){ $ \rho_{5} ( 2350 ) $ }
      \put(40,28){ $ \rho_{3} ( 2250 ) $ }
      \put(32,61){ $ a_{6} ( 2450 ) $ }
      \put(89,60){ $ u \bar{s} $ }
      \put(74,63){ $ J $ }
      \put(148,1){ $ M^{2} $ }
      \put(139,61){ $ J = L + 1 $ }
      \put(139,52){ $ J = L $ }
      \put(139,43){ $ J = L - 1 $ }
      \put(71,15){ $ K^{\ast} ( 892 ) $ }
      \put(79,22){ $ K_{2}^{\ast} ( 1430 ) $ }
      \put(86,8){ $ K_{1} ( 1400 ) $ }
      \put(90,-3){ $ K_{0}^{\ast} ( 1430 ) $ }
      \put(90,31){ $ K_{3}^{\ast} ( 1780 ) $ }
      \put(98,17){ $ K_{2} ( 1770 ) $ }
      \put(108,9){ $ K^{\ast} ( 1680 ) $ }
      \put(100,40){ $ K_{4}^{\ast} ( 2045 ) $ }
      \put(122,30){ $ \longleftarrow $ }
      \put(131.5,28){ $ K_{3} ( 2320 ) $ }
      \put(116,20.8){ $ \rightarrow $ }
      \put(121,19){ $ K_{2}^{\ast} ( 1980 ) $ }
      \put(114,50){ $ K_{5}^{\ast} ( 2380 ) $ }
      \put(132,39.5){ $ \leftarrow $ }
      \put(128,36){ $ K_{4} ( 2500 ) $ }
    \end{picture}
  \end{center}
\caption{$ u \bar{u} $ and $ u \bar{s} $ Regge trajectories}
\label{fig4}
\end{figure}

Both the new results based on $M^2$
(crosses in figs.\ref{fig1}-\ref{fig3} and dashed lines in
fig.\ref{fig4})
and the old ones based on $H_{\rm CM}$
(circlets in figs.\ref{fig1}-\ref{fig3}
and dotted lines in fig.\ref{fig4})
are reported in figures for the parameters discussed in the
introduction.
For the $ l > 0 $ cases masses represent the center of
mass of the multiplets.
In both cases the agreement with the data is on the whole
good, not only for bottonium and charmonium (as in ordinary potential
models), but also the light-light and light-heavy systems. Notice that the
quadratic formulation seems to give a better low angular momentum light-light
spectrum, while perhaps the linear formulation gives 
better Regge trajectories. The Regge trajectories in the quadratic case can 
be improved rising the
value of $\sigma$ at the price, however, of making more difficult the
fitting of
the light-heavy states (perhaps the MAL model is too naive). As we mentioned
the only serious disagreement remains that of the light pseudoscalar
mesons.


\end{document}